# High power, high efficiency, continuous-wave supercontinuum generation using standard telecom fibers

**S Arun, Vishal Choudhury, V Balaswamy, Roopa Prakash and V R Supradeepa**[*]

*Centre for Nano Science and Engineering, Indian Institute of Science, Bangalore, India*
*[*]supradeepa@iisc.ac.in*

**Abstract:** We demonstrate a simple module for octave spanning continuous-wave supercontinuum generation using standard telecom fiber. This module can accept any high power Ytterbium-doped fiber laser as input. The input light is transferred into the anomalous dispersion region of the telecom fiber through a cascade of Raman shifts. A recently proposed Raman laser architecture with distributed feedback efficiently performs these Raman conversions. A spectrum spanning over 1000nm(>1 octave) from 880-1900nm is demonstrated. The average power from the supercontinuum is ~34W with a high conversion efficiency of 44%. Input wavelength agility is demonstrated with similar supercontinua over a wide input wavelength range.

## 1. Introduction

Supercontinuum sources based on optical fibers have gained wide popularity in the past two decades, especially after the demonstration of supercontinuum generation in photonic crystal fibers [1]. Supercontinuum lasers generated using optical fibers provide nearly single mode output with very high brightness, and finds applications in a variety of fields like spectroscopy, OCT, LIDAR and communications [2-4]. In principle, supercontinuum can be generated in any nonlinear medium by pumping very high intensity of light into it. And this can be very effectively done in optical waveguides where high power light can be very tightly confined. This ability of optical fibers to confine high power light over long lengths make them useful as a nonlinear medium for supercontinuum generation. The high peak powers associated with pulsed lasers can generate nonlinear spectral broadening in a medium even at short lengths very easily, leading to supercontinuum generation. However the average power of these pulsed lasers is low when compared with the continuous-wave (CW) counterparts. This manifests as enhanced spectral power density in CW supercontinua which increases its applicability. High power CW supercontinuum generation in optical fibers is initiated by pumping light in the anomalous dispersion region of the fiber near its zero dispersion wavelength (ZDWL). The contributions from nonlinear mechanisms like stimulated Raman scattering (SRS) and four wave mixing (FWM) results in supercontinuum generation [5].

Extensive work has been carried out in photonic crystal fiber based continuous-wave supercontinuum generation by using theoretical simulations and experiments [6-8]. The main attraction of PCF is its high optical nonlinearity and the ability to have ZDWL in the 1micron region where high power Ytterbium doped fiber lasers are easily available. However, limitations in this technology include cost, requirement of free-space optics for input and output coupling and higher fiber loss resulting in reduced efficiency and requirement for better thermal management. These factors hinder its potential for power scaling. Supercontinuum sources based on conventional silica fibers have the advantage that they can be easily fusion spliced with very low loss and do not require any free space optics for coupling of light. This enables an all fiber architecture which can achieve power scaling. Continuous-wave supercontinuum sources based on silica fibers with power levels ranging from few watts to few 10s of watts were demonstrated earlier, mainly using highly nonlinear fibers (HNLF) [9-11]. The nonlinear coefficient of HNLF is an order of magnitude higher than the standard Silica fibers. The zero dispersion wavelength (ZDWL) of these fibers is centered near 1.5um and these systems were pumped using Erbium fiber lasers or Raman lasers that operate near 1.5um [9-12].

The wavelength cutoff of the supercontinuum in the longer wavelength side in optical fibers is limited by the silica absorption near 2um. The total bandwidth of the supercontinuum is dependent on the degenerate and non-degenerate Four-wave mixing (FWM) between the pump near the ZDWL and other wavelengths. The position of the ZDWL affects the shorter wavelength cutoff since the FWM mixing between the pump near the ZDWL and the long wavelength cutoff decides the cutoff in the short-wavelength side. As the ZDWL in HNLF is towards the longer wavelength side (than standard Silica fibers), the shorter wavelength cutoff also moves towards longer wavelengths. Therefore, in order to obtain a wider supercontinuum it is desirable to use optical fibers with lower ZDWL.

The lowest possible ZDWL in conventional Silica fibers is near 1.3um obtained with standard single mode fibers such as SMF28 used in telecom applications. We took advantage of this low ZDWL of telecom fibers and have utilized them to build a simple module which can convert any standard, high power Ytterbium-doped fiber laser into an octave spanning supercontinuum. In this work, we demonstrate a CW supercontinuum pumped by a standard ytterbium doped fiber laser at 1117nm with its spectrum spanning over 1000nm (>1 octave) from 880-1900nm. The average power from the supercontinuum is ~34W with a high conversion efficiency of 44%. We also demonstrate the wavelength agility of the supercontinuum generation module by demonstrating similar supercontinuum spectra and similar power levels for input wavelength varied over a wide wavelength range (>50nm) in the Ytterbium emission window.

## 2. Experiment

The spectral broadening of continuous wave light is initiated when it is pumped near the ZDWL of the optical fiber, which results in breaking up of continuous wave light into pulses due to the onset of modulation instability (MI) [13]. This leads to spectral generation and along with other nonlinear mechanisms like Stimulated Raman Scattering (SRS), Raman Induced Frequency Shift (RIFS) and Four Wave Mixing (FWM), further spectral broadening takes place.

In our experiment, the supercontinuum generation module was constituted of 2 km of standard telecom fiber with it's ZDWL near 1310nm. For high power near 1310nm, we have utilized a recently developed, Ytterbium fiber laser pumped, grating-free, cascaded Raman laser based on distributed feedback [14-18]. The pump wavelength (which is the emission wavelength of Yb laser) in the normal dispersion region undergoes a series of cascaded Raman shifts in the telecom fiber, to longer wavelengths beyond 1.3micron, thereby transferring the power to anomalous dispersion region. In order to achieve efficient cascaded Raman conversion in a single pass architecture it is essential to provide a feedback in the forward direction which will induce preferential forward Raman scattering. This enhances the conversion efficiency and stability [14]. We have used a Raman conversion module based on distributed feedback technique [15, 16] that can provide grating free, wavelength independent feedback [17, 18].

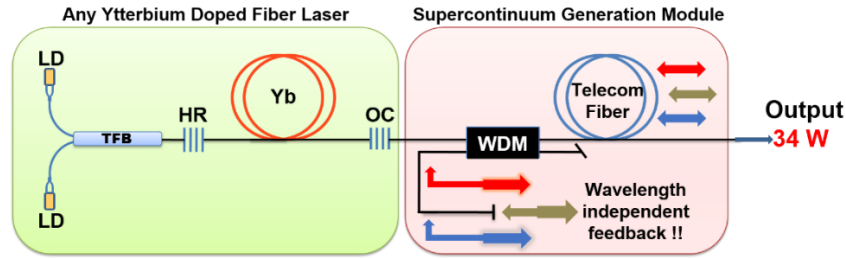

Fig. 1. Architecture for supercontinuum laser generation.

Similar pumping schemes for supercontinuum sources were used earlier with HNLF based supercontinuum in which, cascaded Raman conversions were done till 1.5um, but in a single-pass without the use of feedback enhancement [10]. Such approaches result in comparatively lower efficiencies due to lack of preferential forward scattering and also potential for instabilities due to large backward power into the Ytterbium doped fiber laser pump. These issues are fully avoided in our seeded architecture for efficient, reliable cascaded Raman conversion.

The schematic of our system is shown in Fig 1. A high power Ytterbium(Yb) doped fiber laser operating at 1117nm generating upto 100W of single mode output at full power, was used as the pump laser source. The Yb laser architecture consists of ~20m of Yb doped fiber spliced within a cavity formed by a pair of fiber Bragg gratings (HR and OC) and pumped at 976 nm using laser diodes (LD).The output from the Yb laser is fed into the telecom fiber through a wavelength division multiplexer (WDM) operating between the 1117/1480nm wavelengths. As the pump power at 1117 nm enters the telecom fiber, it undergoes cascaded stimulated Raman scattering (SRS) and a small amount of Raman shifted components are also scattered in the backward direction [19]. A fraction of this backscattered light is cross coupled into the unused input port of the WDM owing to it having different wavelengths. In this port of the WDM, a flat cleave is provided. The flat cleave acts as a glass-air interface which provides ~ 4% Fresnel reflection for the backscattered light which is coupled into the unused input port. The light reflected at the interface propagates in the forward direction which then acts as the seed for the forward cascaded Raman conversion. The choice of 1117/1480nm WDM is not special and any WDM which can effectively separate the pump band (1050-1120nm) from all the higher order Raman stokes components (until the ZDWL) can be used (for example 1060/1310). In order to understand the importance of providing seeding for Raman conversions, we made an angle cleave (8.5 degree) instead of flat cleave at the unused input port of WDM. We then observed the onset of temporal oscillations at the output as soon as we started increasing the power which is a characteristic feature of Raman based instabilities [20]. This prevented any higher order Raman conversion. This seeded Raman conversion module improves the Raman conversion efficiency by ensuring preferential forward scattering and avoid any laser instabilities. Cascaded Raman conversion occurs until the light moves into the anomalous dispersion region of the fiber at which point modulation instability seeded supercontinuum generation occurs.

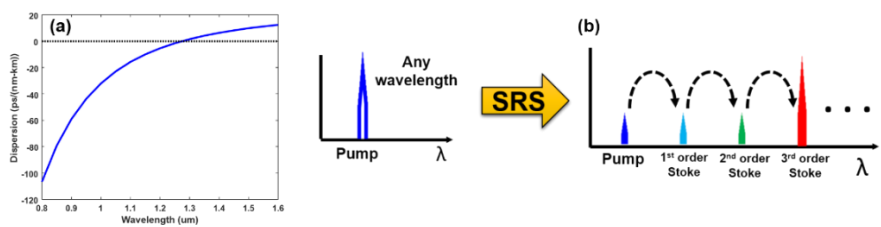

Fig. 2. (a) Dispersion profile of telecom fiber (b) schematic of SRS along the fiber.

In our experiment 76W at 1117nm was sent through the supercontinuum generation module. The dispersion profile of silica fiber is shown in Fig 2 (a). The schematic for growth of higher order Raman stokes through SRS, with increase in power, starting from 1117 nm pump inside the telecom fiber, is shown in Fig 2 (b). By undergoing cascaded Raman conversion, the power from the 1117 nm pump is transferred to the anomalous dispersion region (at 1310nm), after 3 Raman stokes conversions. Fig 3 shows the growth of the spectrum with increase in pump power, measured using an OSA. The growth of higher order Raman stokes in the normal dispersion region with increase in pump power is shown in Fig 3(a) and 3(b). Once the 1310 nm stokes starts growing, CW light breaks into pulses because of MI and this results in spectral broadening as shown in Fig 3(c). When the power is increased further, other nonlinearities extend the spectrum both in shorter and longer wavelength sides, as shown in Fig. 3(d).

While SRS and RIFS extends the spectrum towards the longer wavelength side, FWM and dispersive wave generation mainly extends the spectrum towards the shorter wavelength region [13]. Silica attenuation becomes very high (>10 dB/km) above 1900 nm and this starts terminating the spectrum in the longer wavelength region to around 1900 nm. The parametric FWM between the pump at ZDWL and the longer wavelength components generates the spectrum at shorter wavelength side and the extent of this shorter wavelength cutoff is limited by the longer wavelength cutoff. On the positive side, the presence of Raman stokes lines in the normal dispersion region due to the distributed pumping scheme, undergoes FWM with soliton like pulses in the anomalous dispersion region. This helps in the growth of the dispersive waves better and results in a much smoother spectrum in the normal dispersion region [21].

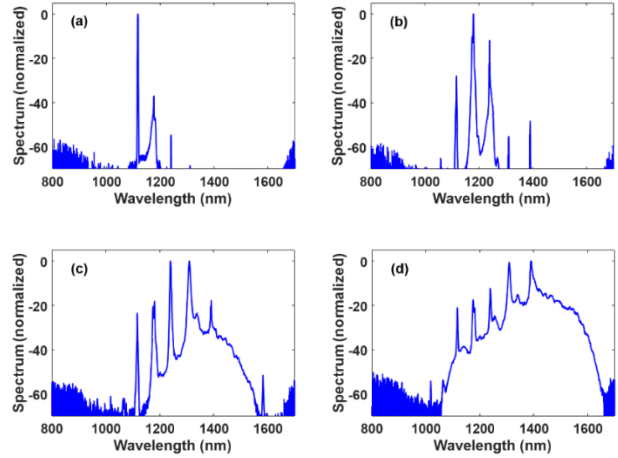

Fig. 3. Supercontinuum evolution at different output powers (a) 3W (b) 8W (c) 11W (d) 15W.

## 3. Results

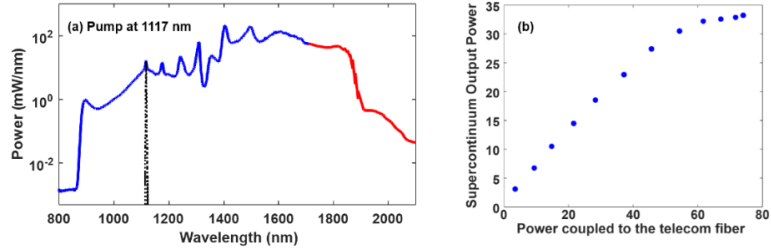

Fig. 4. Full spectra for 1117nm pumping at full power (pump location shown in dotted lines, blue part of spectra captured by OSA, red part of spectra captured by mid-IR spectrometer) (b) Supercontinuum output power vs. input power coupled to the telecom fiber.

The supercontinuum spectrum spans from 880 nm to 1900 nm within a 20-dB bandwidth for an input power of 76 W, spectrum is shown in Fig 4 (a). The total average (CW) power within the supercontinuum is >34 W with a power spectral density of at least 1mW/nm from 880-1380 nm and >50mW/nm for 1380-1900 nm. The output power from the supercontinuum increases linearly with the pump power till 61 W of pump power, as shown in Fig 4 (b). And with further increase in pump power the supercontinuum power saturates, because the spectrum has grown full in the longer wavelength side and further power transfer to longer wavelengths will be attenuated by the silica fiber losses. This leads to the saturation of SC power when pump power is increased further.

We used the standard Yokogawa AQ 6370 OSA for the spectral measurements which was limited to 1700 nm. For wavelengths beyond this, we used an in-house built mid-IR spectrometer with ~6nm resolution. In order to demonstrate input wavelength agility of the supercontinuum module, we pumped it with different pump wavelengths in the Yb emission band using a tunable Yb laser operating between 1060-1100nm [22]. Fig 5 shows the supercontinuum spectrum for 1073, 1079, 1085 and 1088 nm pump wavelengths. The bandwidth and shape of the spectrum is nearly the same for all pump wavelengths. The output power was measured to be similar in all cases (>30W). These results together with the 1117nm pumping results shows that we can use this architecture with any Yb laser, operating at any wavelength in its emission band.

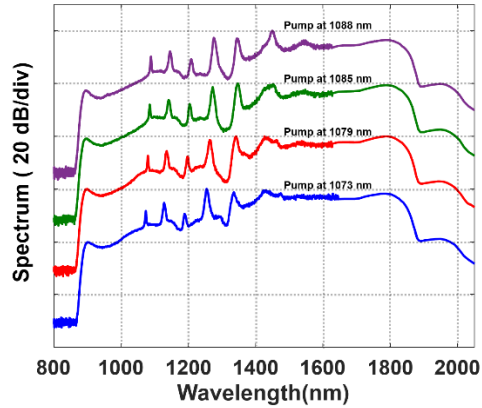

Fig. 5. Supercontinuum spectra for different pump wavelengths

## 4. Summary

We have demonstrated a simple, all standard fiber module to convert any high power Yb laser into an octave spanning supercontinuum in a cost-effective manner by using telecom fiber as the nonlinear medium. In this work, the supercontinuum generates around 34W of CW power, limited by pump power, over a bandwidth of ~1000 nm extending from 880-1900 nm with a substantial power spectral density (>1mW/nm from 880-1380 nm and >50mW/nm for 1380-1900 nm). A high conversion efficiency of ~44% was demonstrated. A standard, high power Yb laser was used as the pump laser source and the pump power was wavelength shifted to the anomalous dispersion region with high efficiency through a new, grating-free, cascaded Raman convertor based on distributed feedback. We also demonstrated the wavelength agility of the module by demonstrating similar supercontinuum spectra and power levels for input wavelength varied over a wide wavelength range (>50nm) in the Ytterbium emission window. Further power scaling can be achieved with increased input power and an appropriate modification of the SMF fiber length. The proposed module promises to be an exciting, power scalable technique to convert standard Yb fiber lasers into octave spanning supercontinua.

## Funding